\documentclass{article}
\usepackage{amsmath, amssymb, amsfonts}
\title{On the conformally flat Rindler-like geometry} 
\author{Hristu Culetu, \\Ovidius University, Dept.of Physics, \\B-dul Mamaia 124, 900527 Constanta, Romania, \\e-mail : hculetu@yahoo.com}

\begin{document}
\numberwithin{equation}{section}
\pagenumbering{arabic}
\maketitle
\newcommand{\fv}{\boldsymbol{f}}
\newcommand{\tv}{\boldsymbol{t}}
\newcommand{\gv}{\boldsymbol{g}}
\newcommand{\OV}{\boldsymbol{O}}
\newcommand{\wv}{\boldsymbol{w}}
\newcommand{\WV}{\boldsymbol{W}}
\newcommand{\NV}{\boldsymbol{N}}
\newcommand{\hv}{\boldsymbol{h}}
\newcommand{\yv}{\boldsymbol{y}}
\newcommand{\RE}{\textrm{Re}}
\newcommand{\IM}{\textrm{Im}}
\newcommand{\rot}{\textrm{rot}}
\newcommand{\dv}{\boldsymbol{d}}
\newcommand{\grad}{\textrm{grad}}
\newcommand{\Tr}{\textrm{Tr}}
\newcommand{\ua}{\uparrow}
\newcommand{\da}{\downarrow}
\newcommand{\ct}{\textrm{const}}
\newcommand{\xv}{\boldsymbol{x}}
\newcommand{\mv}{\boldsymbol{m}}
\newcommand{\rv}{\boldsymbol{r}}
\newcommand{\kv}{\boldsymbol{k}}
\newcommand{\VE}{\boldsymbol{V}}
\newcommand{\sv}{\boldsymbol{s}}
\newcommand{\RV}{\boldsymbol{R}}
\newcommand{\pv}{\boldsymbol{p}}
\newcommand{\PV}{\boldsymbol{P}}
\newcommand{\EV}{\boldsymbol{E}}
\newcommand{\DV}{\boldsymbol{D}}
\newcommand{\BV}{\boldsymbol{B}}
\newcommand{\HV}{\boldsymbol{H}}
\newcommand{\MV}{\boldsymbol{M}}
\newcommand{\be}{\begin{equation}}
\newcommand{\ee}{\end{equation}}
\newcommand{\ba}{\begin{eqnarray}}
\newcommand{\ea}{\end{eqnarray}}
\newcommand{\bq}{\begin{eqnarray*}}
\newcommand{\eq}{\end{eqnarray*}}
\newcommand{\pa}{\partial}
\newcommand{\f}{\frac}
\newcommand{\FV}{\boldsymbol{F}}
\newcommand{\ve}{\boldsymbol{v}}
\newcommand{\AV}{\boldsymbol{A}}
\newcommand{\jv}{\boldsymbol{j}}
\newcommand{\LV}{\boldsymbol{L}}
\newcommand{\SV}{\boldsymbol{S}}
\newcommand{\av}{\boldsymbol{a}}
\newcommand{\qv}{\boldsymbol{q}}
\newcommand{\QV}{\boldsymbol{Q}}
\newcommand{\ev}{\boldsymbol{e}}
\newcommand{\uv}{\boldsymbol{u}}
\newcommand{\KV}{\boldsymbol{K}}
\newcommand{\ro}{\boldsymbol{\rho}}
\newcommand{\si}{\boldsymbol{\sigma}}
\newcommand{\thv}{\boldsymbol{\theta}}
\newcommand{\bv}{\boldsymbol{b}}
\newcommand{\JV}{\boldsymbol{J}}
\newcommand{\nv}{\boldsymbol{n}}
\newcommand{\lv}{\boldsymbol{l}}
\newcommand{\om}{\boldsymbol{\omega}}
\newcommand{\Om}{\boldsymbol{\Omega}}
\newcommand{\Piv}{\boldsymbol{\Pi}}
\newcommand{\UV}{\boldsymbol{U}}
\newcommand{\iv}{\boldsymbol{i}}
\newcommand{\nuv}{\boldsymbol{\nu}}
\newcommand{\muv}{\boldsymbol{\mu}}
\newcommand{\lm}{\boldsymbol{\lambda}}
\newcommand{\Lm}{\boldsymbol{\Lambda}}
\newcommand{\opsi}{\overline{\psi}}
\renewcommand{\tan}{\textrm{tg}}
\renewcommand{\cot}{\textrm{ctg}}
\renewcommand{\sinh}{\textrm{sh}}
\renewcommand{\cosh}{\textrm{ch}}
\renewcommand{\tanh}{\textrm{th}}
\renewcommand{\coth}{\textrm{cth}}

\begin{abstract}
 The time dependent conformally-flat spherical Rindler spacetime is investigated. The geometry has an apparent horizon that coincides with the causal horizon. The scalar acceleration of a static observer is constant and equals to the acceleration $g$ from the static expression of the Rindler-like metric. The timelike radial geodesics are computed and proves to be hyperbolae when a specific choice of the constants of integration is operated.
\end{abstract}

\textbf{Introduction}

 The matter content of the Universe consists of two basic components: dark matter (DM) and dark energy (DE), with ordinary matter playing a minor role. The nature and composition of DM and DE is not at all understood. The entire motivation of their existence is based on their validity at all distance scales \cite{PM1, DG1}. Starting with a conformally invariant Lagrangean, Mannheim obtained a spherically-symmetric vacuum solution of 4-th order Einstein's equation with no cosmological term, with $-g_{tt} = g^{-1}_{rr} = 1 - (2M/r) + \gamma_{0}r$, where $\gamma_{0}$ is a constant and $M$ is the central mass. The metric is no longer asymptotically flat. In Mannheim's view, the linearly rising potential $\gamma_{0}r$ shows that a local matter distribution can actually have a global effect at infinity and gravitational theories become global.
 
 Kiselev \cite{VVK} investigated black holes (BH) surrounded by quintessential matter whose energy-momentum tensor obeys the conditions of additivity and linearity. Apart from the Schwarzschild term $(-2M/r)$, the metric contains a term proportional to $r^{-(3\omega_{q}+1)}$, where $r$ is the radial coordinate and $\omega_{q}$ is obtained from the equation of state for the quintessence: $p_{q} = \omega_{q}\epsilon_{q}, (\epsilon_{q}$ and $p_{q}$ represents the energy density and, respectively, the quintessential pressure). The value $\omega_{q} = -2/3$ leads Kiselev to a term in the metric of the form $(-ar)$, with $a$ a constant. It has the same physical meaning as the Rindler-like term from Mannheim's and Grumiller's papers (see also \cite{HC6}). Without the mass term the spacetime corresponds to the self-gravitating quintessence (free quintessence) \cite{VVK}. It is worth to remark that Kiselev obtained his solution for a BH surrounded by quintessence, using a complicated isotropic average over the angles, applied upon the quintessential stress tensor. However, his Rindler-type metric (23) from \cite{VVK} has been previously obtained by Mannheim \cite{PM1} and studied later in \cite{DG1, HC5} in connection with the flat rotation curves of spiral galaxies.
 
 Fernando \cite{SF} studied the quintessence matter, focusing on the special choice $\omega_{q} = -2/3$. Then she investigated the null geodesics in that special geometry, both for radial and nonzero angular momentum geodesics. \\  
 Throughout the paper we consider geometrical units $G = c = \hbar = 1$.\\
 
 \textbf{Time dependent Rindler-like metric}
         
We used in \cite{HC5} the Mannheim - Grumiller metric without the mass term
\begin{equation}
 ds^{2} = -(1 - 2gr) dt^{2} + (1 - 2gr)^{-1} dr^{2} + r^{2} d \Omega^{2},
\label{1}
\end{equation}
where $g$ is the ''Rindler'' acceleration and $d \Omega^{2}$ stands for the metric on the unit 2 - sphere, inside a relativistic star. When it is applied in the black hole interior  \cite{HC6}, the Rindler acceleration equals the surface gravity, i.e. $g = 1/4M$.
 
To be a solution of Einstein's equations $G_{ab} = 8\pi T_{ab}$, one shows that a stress tensor is needed on its r.h.s. , namely
\begin{equation}
 T_{~t}^{t} = - \epsilon = - \frac{g}{2 \pi r},~~~p_{r} = T_{~r}^{r} = - \epsilon,~~~T^{\theta}_{~\theta} = T^{\phi}_{~\phi} = p_{\bot} = \frac{1}{2} p_{r}
 \label{2}
 \end{equation}
 where $\epsilon$ is the energy density of the anisotropic fluid, $p_{r}$ is its radial pressure and $p_{\bot}$ are the tangential pressures. It is worth noting that the anisotropic fluid is comoving with the accelerated observer (the stress tensor is diagonal). We chose $g > 0$ in order for the metric (1) to possess a Rindler horizon. We found \cite{HC5} that, in the metric (1)
  \begin{equation}
 R_{~a}^{b} = diag(\frac{2g}{r}, \frac{2g}{r}, \frac{4g}{r}, \frac{4g}{r}),~~R_{~a}^{a} = \frac{12g}{r}, ~~R_{~~~~abcd}^{abcd} = \frac{32g^{2}}{r^{2}}
\label{3}
\end{equation}
and the components of the stress tensor (2) can be extracted from the general expression of the energy - momentum tensor for an anisotropic fluid 
 \begin{equation}
T_{a}^{b} = (\epsilon + p_{\bot})u_{a} u^{b} + p_{\bot} \delta_{a}^{b}+ (p_{r} - p_{\bot}) s_{a} s^{b}.
\label{4}
\end{equation}
 The Latin indices run from $0$ to $3$ in the order $(t, r, \theta, \phi)$. $u^{a}$ is the timelike velocity vector of the fluid and $s^{a}$ is spacelike , on the direction of anisotropy, with $s^{a} u_{a} = 0$. They are given by
 \begin{equation}
 u^{a} = (\frac{1}{\sqrt{1 - 2gr}}, 0, 0, 0)~,~~~s^{a} = (0, \sqrt{1 - 2gr}, 0, 0) 
\label{5}
\end{equation}
Even though the Ricci tensor from (3) has non zero components, the Weyl tensor vanishes in the geometry (1) and, therefore, it may be written as a conformally flat spacetime. Indeed, Kiselev \cite{VVK} (see also \cite{HC6}) brought the metric (1) in a conformally flat form. We display, for completeness, his most important steps.

With the help of the coordinate transformation  $\chi = (-1/2) ln(1 - 2gr)$, (1) may be expressed as 
\begin{equation}
ds^{2} = \frac{1}{g^{2}} e^{-2\chi} (- g^{2} dt^{2} + d\chi^{2} + sinh^{2}\chi d \Omega^{2}).
\label{6}
\end{equation}
 Kiselev further makes the Fock transformation from the hyperbolic coordinates to the flat ones
\begin{equation}
\tau = e^{g t} cosh\chi,~~~\rho = e^{g t} sinh\chi,
\label{7}
\end{equation}
to find
\begin{equation}
ds^{2} = \frac{1}{g^{2}(\tau + \rho)^{2}} (- d\tau^{2} + d\rho^{2} + \rho^{2} d \Omega^{2}),
\label{8}
\end{equation}
which has the desired conformally flat structure. We have above $\chi > 0,~\tau > 0$ and $0 < \rho <\tau$. The last equation reveals $\rho = 0$ as a singular point (it corresponds to $r = 0$ in the original metric). In addition, the condition $\rho + \tau \rightarrow \infty$ is equivalent to $r \rightarrow 1/2g$ in (1) (we note that (8) is not spatially homogeneous due to the $\rho$-dependence of the conformal factor). Hence, in contrast to the conformally flat de Sitter universe the geometry (8) does not allow a FRW type evolution \cite{VVK}. \\

\textbf{Apparent horizon and the Misner - Sharp energy}

Since the metric (1) is time-dependent, the usual event horizon is not useful to be studied (there is no a timelike Killing vector whose modulus would vanish on the horizon). We therefore have to study the apparent horizon (AH) obtained from the equation \cite{HKS} 
  \begin{equation}
 g^{ab} \nabla_{a}R ~\nabla_{b}R = 0,
 \label{9}
 \end{equation}
 where $R(\rho,\tau) = \rho/g(\tau + \rho)$ is the areal radius. Solving (9), one obtains
 \begin{equation}
 \rho^{2} - \tau^{2} = 0.
 \label{10}
 \end{equation}
Keeping in mind that we have $\rho > 0,~\tau > 0$, Eq. (10) yields
 \begin{equation}
 \rho_{AH} = \tau,
 \label{11}
 \end{equation}
namely the apparent horizon actually represents the null radial geodesic (causal horizon) for the spacetime (1).

The Misner - Sharp (MS) energy $E$ contained by a sphere with radius equal to the areal (physical) radius is obtained from \cite{NY, HPFT}
  \begin{equation}
  1 - \frac{2E(t,r)}{R} = g^{ab} \nabla_{a}R ~\nabla_{b}R,
 \label{12}
 \end{equation}
whence
  \begin{equation}
  E = \frac{\rho^{2}}{g(\tau + \rho)^{2}} = \frac{1}{g(1 + \frac{\tau}{\rho})^{2}}.
 \label{13}
 \end{equation}
In other words, $E$ depends only on the ratio $\tau/\rho$. The same is valid for the energy density $\epsilon$ from (2). When $\epsilon$ is written in terms of the coordinates (8), we have, indeed
  \begin{equation}
  \epsilon = -p_{r} = 4g^{2} (1 + \frac{\tau}{\rho}).
 \label{14}
 \end{equation}
We could convince ourselves that the Ricci and Kretschmann scalars from Eqs. (3) share a similar property. It is interesting to notice that the MS energy acquires the value $E_{AH} = 1/4g$ on the apparent horizon. A similar expression has been obtained in \cite{HC4} for the Rindler horizon energy in the standard (flat) Rindler metric. It is an easy task to check that $E$ may be put in the form \cite{BI, HC3}
  \begin{equation}
  E = \frac{4 \pi}{3}R^{3}(\epsilon - p_{r} + p_{\bot}),
 \label{15}
 \end{equation}
  where the MS mass $E$ represents the Ricci part $E_{R}$ since the Weyl component $E_{W}$ is vanishing due to the conformal flatness of the metric. \\

\textbf{Congruence of static observers}

Let us consider a static ($\rho = const.)$ observer with the velocity vector field
  \begin{equation}
  u^{a} = (g(\tau + \rho), 0, 0, 0)
 \label{16}
 \end{equation}
with $u^{a}u_{a} = -1$. The static observer is not geodesic and its acceleration $a^{b} = u^{a}\nabla_{a}u^{b}$ has the components
  \begin{equation}
  a^{b} = (0, -g^{2}(\tau + \rho), 0, 0)
 \label{17}
 \end{equation}
with a constant modulus $\sqrt{a^{b}a_{b}} = g$. The radial component in (17) is negative (towards the origin of coordinates) because $a^{\rho}$ is the acceleration needed to maintain our observer at fixed $\rho$ and the gravitational field is repulsive due to the negative pressures (as for dark energy).
In contrast, the invariant normal acceleration on the surface of constant $\rho$, given by $a^{b}n_{b} = -g$, is constant. The normal $n^{b}$ to $\rho$ = const. surface has only one nonzero components, $n^{\rho} = g(\tau + \rho)$. 

While the shear and vorticity tensors of the congruence are vanishing, one finds that the expansion scalar reads
  \begin{equation}
 \Theta \equiv \nabla_{a}u^{a} = -3g,~~~\dot{\Theta} \equiv u^{a}\nabla_{a} \Theta = 0. 
 \label{18}
 \end{equation}
The fact that $\Theta < 0$ is rooted from the repulsive character of the field and from the static feature of our observer.\\

\textbf{Timelike radial geodesics}

The geodesic trajectories for the metric (1) have been studied in \cite{HC6}. We fould that the radial timelike geodesics appears as
  \begin{equation}
  r(t) = \frac{1}{2g} \left(1 - \frac{E^{2}}{cosh^{2}gt}\right) ,
\label{19}
\end{equation}
under suitable initial conditions ($0 < E < 1$ is a constant related to observer's energy and $t\in(-\infty, \infty)$). Notice that there is a $r_{min} = (1 - E^{2})/2g$, as if the observer (a test particle) were rejected by a central core, due to the singularity at the origin $r = 0$. 

Regarding the null radial geodesics one obtains from (1)
  \begin{equation}
  r(t) = \frac{1}{2g} \left(1 - e^{- 2g|t|} \right) ,~~~t \in(- \infty, \infty),
\label{20}
\end{equation}
 Hence, once the light signal reaches the central singularity at t = 0, it bounces there and then becomes outgoing, reaching the horizon r = 1/2g when $t \rightarrow \infty$.
 
Let us now study how the timelike geodesics look like in the coordinates ($\tau, \rho$), in terms of which the spacetime is given by (8). Starting with the standard
geodesic equations
  \begin{equation}
 \frac{d^{2}x^{a}}{ds^{2}} + \Gamma^{a}_{bc} \frac{dx^{b}}{ds} \frac{dx^{c}}{ds} = 0
\label{21}
\end{equation}
one obtains for the $\tau-$ and $\rho-$ components of (21)
  \begin{equation}
\frac{d^{2}\tau}{ds^{2}} - \frac{1}{\tau + \rho} \left[\frac{d}{ds} \left(\tau + \rho\right)\right]^{2} = 0  
\label{22}
\end{equation}
and, respectively
  \begin{equation}
\frac{d^{2}\rho}{ds^{2}} - \frac{1}{\tau + \rho} \left[\frac{d}{ds} \left(\tau + \rho\right)\right]^{2} = 0  
\label{23}
\end{equation}
Subtracting and then adding the two equations, we have
  \begin{equation}
\frac{d^{2}}{ds^{2}}\left(\tau - \rho\right) = 0,~~~ \frac{d^{2}}{ds^{2}}\left(\tau + \rho\right) - \frac{2}{\tau + \rho} \left[\frac{d}{ds} \left(\tau + \rho\right)\right]^{2} = 0
\label{24}
\end{equation}
The above expressions give us 
  \begin{equation}
  \tau - \rho = \alpha s - \beta~:~~~\tau + \rho = - \frac{\gamma}{s + \delta}
\label{25}
\end{equation}
where $\alpha, \beta, \gamma, \delta$ are constants of integration. We get rid of $s$ in (25) and obtain
  \begin{equation}
  \tau - \rho = \alpha \left(- \frac{\gamma}{\tau + \rho} - \delta\right) + \beta
\label{26}
\end{equation}
whence
  \begin{equation}
  \rho^{2} - \tau^{2} + \left(\beta - \alpha \delta\right)\left(\tau + \rho\right) = \alpha \gamma,
\label{27}
\end{equation}
that is an arc of hyperbola for $\rho(\tau)$ (we chose $\alpha \gamma > 0,~\beta - \alpha \delta > 0$ because $\rho < \tau$ always).

Eq. (27) can be written as
  \begin{equation}
  \left(\bar{\rho} + \kappa\right)^{2} - \left(\bar{\tau} - \kappa\right)^{2} = 1
\label{28}
\end{equation}
where $\bar{\rho} \equiv \rho/\sqrt{\alpha \gamma},~\bar{\tau} \equiv \tau/\sqrt{\alpha \gamma}$ and $\left(\beta - \alpha \delta\right)/\alpha \gamma \equiv 2\kappa > 0$. Even though $\bar{\rho} < \bar{\tau}$, Eq. (28) may be obeyed provided $\bar{\tau} - \bar{\rho} < 2\kappa$.

A similar hyperbola like (28) would have been obtained directly from (19) by means of the transformations (7) and under suitable initial conditions. One indeed obtains
  \begin{equation}
  \left(\rho + E\right)^{2} - \left(\tau - E\right)^{2} = 1
\label{29}
\end{equation}
with $\tau - \rho < 2E$. Now, taking $\alpha \gamma = 1$ in (28), the last two equations are identical provided $\kappa = E$. In that case, the minimum value for $\rho$ is obtained from (29), keeping in mind that $\rho \leq \tau$ always. We get $\rho_{min} = 1/4E$. $\rho(\tau)$ has only one asymptote given by $\rho = \tau - 2E$.

It is worth noting that $\tau$ and $\rho$ are dimensionless. If we define $\rho = g\bar{r},~\tau = g\bar{t}$, $\bar{r}$ and $\bar{t}$ have dimension of distance and (29) becomes
  \begin{equation}
  \left(\bar{r} + l\right)^{2} - \left(\bar{t} - l\right)^{2} = \frac{1}{g^{2}}
\label{30}
\end{equation}
where $l \equiv E/g$. From (30) it is clear that the constant $g$ from the geometry (1) represents the acceleration of the geodesic observer.\\

\textbf{Conclusions}

While Kiselev introduced a quintessential matter as a source to get a Rindler-type metric as a solution of Einstein's equation, we have used an anisotropic fluid stress tensor on its r.h.s. to reach the same solution. The time dependent conformal version of the metric is studied in detail in this paper. The $MS$ energy is calculated and has only a nonzero Ricci term (the Weyl term vanishes due to the conformal flatness of the metric). It is interesting that the apparent horizon represents a null radial geodesic (the causal horizon). The timelike radial geodesics $\rho(\tau)$ are analyzed and seems to be a part of a hyperbola. There is a minimum value of the spatial coordinate, $\rho_{min} = 1/4E$, obtained from the condition $\rho \leq \tau$.

\end{document}